\documentstyle[prd,aps,preprint]{revtex}
\newcommand{\be}{\begin{equation}}
\newcommand{\ee}{\end{equation}}
\newcommand{\bea}{\begin{eqnarray}}
\newcommand{\eea}{\end{eqnarray}}

\newcommand{\bei}{\begin{itemize}}
\newcommand{\eei}{\end{itemize}}
\newcommand{\sla}[1]{#1\!\!\!\slash}

\newcommand{\xp}{x^\prime}
\newcommand{\yp}{y^\prime}
\newcommand{\kp}{k^\prime}
\newcommand{\td}{r^2}

\newcommand{\rhop}{\rho_+}
\newcommand{\rhom}{\rho_-}
\newcommand{\rhod}{\rho^2}

\newcommand{\srho}{\sqrt{\rho^2-1}}

\begin{document}
%\hfill{NCKU-HEP-99-06}

\author{Hsien-Hung Shih\footnote{hhshih@phys.sinica.edu.tw} and
Shih-Chang Lee\footnote{phsclee@ccvax.sinica.edu.tw}
\\
{\small Institute of Physics, Academia Sinica, Taipei, Taiwan 105,
Republic of China}
\\
Hsiang-nan Li\footnote{hnli@mail.ncku.edu.tw}
\\
{\small Department of Physics, National Cheng-Kung University,
Tainan, Taiwan 701, Republic of China} }

\title{Applicability of perturbative QCD to $\Lambda_b \to \Lambda_c$
decays}

\date{\today}

\maketitle

\begin{abstract}
We develop perturbative QCD factorization theorem for the semileptonic
heavy baryon decay $\Lambda_b \to \Lambda_c l\bar{\nu}$, whose form factors 
are expressed as the convolutions of hard $b$ quark decay amplitudes with
universal $\Lambda_b$ and $\Lambda_c$ baryon wave functions. Large
logarithmic corrections are organized to all orders by the Sudakov
resummation, which renders perturbative expansions more reliable. It is
observed that perturbative QCD is applicable to $\Lambda_b \to \Lambda_c$
decays for velocity transfer greater than 1.2. Under requirement of heavy
quark symmetry, we predict the branching ratio
$B(\Lambda_b \to \Lambda_c l{\bar\nu})\sim 2\%$, and determine the
$\Lambda_b$ and $\Lambda_c$ baryon wave functions.

\end{abstract}
\vskip 0.5cm
PACS numbers: 12.38.Bx, 12.38.Cy, 13.30.Ce, 12.39.Fe, 11.30.Rd

\section{INTRODUCTION}

Analyses of exclusive heavy hadron decays are a challenging subject because
of their complicated QCD dynamics. Recently, we have proposed a rigorous
theory for these processes based on perturbative QCD (PQCD) factorization
theorems \cite{LY1,CL}. In this approach heavy hadron decay rates are
expressed as convolutions of hard heavy quark decay amplitudes with heavy
hadron wave functions. The former are calculable in perturbation theory, if
processes involve large momentum transfer. The latter, absorbing
nonperturbative dynamics of processes, must be obtained by means outside the
PQCD regime. Since wave functions are universal, they can be
determined once for all, and then employed to make predictions for other
processes containing the same hadrons. With this prescription for
nonperturbative wave functions, PQCD factorization theorems possess a
predictive power.

For semileptonic decays, the PQCD approach complements heavy quark symmetry
in studies of heavy hadron transition form factors \cite{L1}. Heavy quark
symmetry determines the normalization of transition form factors at zero
recoil of final-state heavy hadrons, up to power corrections in $1/M$, $M$
being the heavy quark mass, and up to perturbative corrections in the
coupling constant $\alpha_s$. While PQCD is appropriate for fast recoil,
the region with large energy release, and gives a dependence of transition
form factors on velocity transfer. For nonleptonic decays, PQCD is
a more systematic approach compared with the phenomenological
Bauer-Stech-Wirbel (BSW) model \cite{BSW}. In PQCD factorization theorems
contributions to nonleptonic decay rates characterized by different scales
are carefully absorbed into different subprocesses, among which
renormalization-group (RG) evolutions are constructed \cite{CL}, leading to
a scale and scheme independent, gauge invariant and infrared finite theory
\cite{CLY}. Not only factorizable but nonfactorizable contributions can be
evaluated \cite{WYL}. The BSW model considers only factorizable
contributions: two fitting parameters $a_1$ and $a_2$ are associated with
external and internal $W$-emission form factors, respectively.
Nonfactorizable contributions must be included as additional parameters
\cite{C}.

The above PQCD formalism has been applied to heavy meson decays successfully.
It is then natural to extend the formalism to more complicated heavy baryon
decays. In \cite{SLL} we have developed factorization theorem for the
semileptonic decay $\Lambda_b\to pl\bar{\nu}$, in which Sudakov resummation
of double logarithmic corrections to the $\Lambda_b$ baryon wave function
was included, and a full set of diagrams for the hard $b$ quark decay
amplitudes was calculated. This is an analysis more complete than the work
in the literature \cite{RA}. On the other hand, $b$-baryons have been
observed in experiments at LEP and at the Tevatron. Masses and decay widths
of the lightest $b$-baryons, as compared with theoretical predictions, have
stimulated many interesting discussions and investigations
\cite{Li1,Li2,NS,Alt,HYC1}.
When Run II of the Tevatron comes up with a vertex trigger employed, it
will be expected to collect millions of $b$-baryon events. Therefore,
an intensive study of exclusive heavy baryon decays is urgent.

Exclusive heavy baryon decays are dominated by $b\to c$ modes.
In this paper we shall develop factorization theorem for the semileptonic
decay $\Lambda_b\to \Lambda_cl\bar{\nu}$, and locate the kinematic region
where PQCD is applicable. It will be shown that PQCD predictions for the
involved transition form factors are reliable at fast recoil of the
$\Lambda_c$ baryon with velocity transfer greater than 1.2. Under 
requirement of heavy quark symmetry, we predict
the branching ratio $B(\Lambda_b \to \Lambda_c l{\bar\nu})\sim 2\%$.
We shall also determine the unknown parameters in the $\Lambda_b$ and
$\Lambda_c$ baryon wave functions, which can be employed to study
nonleptonic $\Lambda_b$ baryon decays because of the universality.

In Sec.~II we develop factorization theorem for the semileptonic decay
$\Lambda_b \to \Lambda_c l{\bar\nu}$. Sudakov resummation of double
logarithmic corrections to the process is performed. The factorization
formulas for the involved heavy baryon transition form factors and their
numerical results are presented in Sec.~III and in Sec.~IV, respectively.
Section V is the conclusion.

\section{FACTORIZATION THEOREM}

The amplitude for the semileptonic decay $\Lambda_b\to \Lambda_cl\bar{\nu}$ 
is written as
\bea
{\cal M} = \frac{G_F}{\sqrt{2}}V_{cb}\bar{l}\gamma^{\mu} (1-\gamma_5) \nu_l 
\langle \Lambda_c(p^\prime)|\bar{c}\gamma_{\mu}(1-\gamma_5)b
|\Lambda_b(p)\rangle \;,
\eea
where $G_F$ is the Fermi coupling constant, $V_{cb}$ is the
Cabibbo-Kobayashi-Maskawa (CKM) matrix element, $p$ and $p^\prime$ are the
$\Lambda_b$ and $\Lambda_c$ baryon momenta, respectively. All QCD dynamics
is contained in the hadronic matrix element
\bea
{\cal M}_{\mu} &\equiv& \langle \Lambda_c(p^\prime)|\bar{c}\gamma_{\mu} 
(1-\gamma_5)b| \Lambda_b(p)\rangle\;,
\nonumber\\
&=&\bar{\Lambda}_c(p') [ f_1(q^2)\gamma^\mu
-if_2(q^2)\sigma^{\mu\nu}q_\nu+f_3(q^2)q^\mu ]\Lambda_b(p)
\nonumber \\
& & +\bar{\Lambda}_c(p')[ g_1(q^2)\gamma^\mu\gamma_5
-ig_2(q^2)\sigma^{\mu\nu}\gamma_5q_\nu
+g_3(q^2)\gamma_5q^\mu ]\Lambda_b(p)\;.
\eea
In the second expression ${\cal M}_\mu$ has been expressed in terms of six
form factors $f_i$ and $g_i$, where $\Lambda_b(p)$ and $\Lambda_c(p')$ are
the $\Lambda_b$ and $\Lambda_c$ baryon spinors,
respectively, and the variable $q$ denotes $q=p-p'$.
In the case of massless leptons with
\begin{eqnarray}
q_\mu \bar{l} \gamma^\mu(1-\gamma_5) \nu_l= 0 \;,
\end{eqnarray}
the form factors $f_3$ and $g_3$ do not contribute. Since the contributions
from $f_2$ and $g_2$ are small, we shall concentrate on $f_1$ and $g_1$
in the present work.

The idea of PQCD factorization theorems is to sort out nonperturbative
dynamics involved in QCD processes and factorize it into hadron wave
functions. Nonperturbative dynamics is reflected by infrared divergences in
radiative corrections to quark-level amplitudes in
perturbation theory. The construction of factorization theorem for
the decay $\Lambda_b\to \Lambda_cl\bar{\nu}$ is basically similar to that
for the decay $\Lambda_b\to pl\bar{\nu}$ in \cite{SLL}. The lowest-order
diagrams for $b\to c$ decays are shown in Fig.~1, where two hard gluons
attach the three incoming and outgoing quarks in all possible ways. We then
investigate infrared divergences from radiative corrections to these
diagrams. Small transverse momenta $k_T$ are associated with the valence
quarks, such that they are off mass shell a bit. The transverse momenta
$k_T$ serve as a factorization scale, below which dynamics is regarded as
being nonperturbative, and absorbed into $\Lambda_b$ and $\Lambda_c$ baryon
wave functions, and above which perturbation theory is reliable, and
radiative corrections are absorbed into hard $b\to c$ decay amplitudes.

Infrared divergences from radiative corrections are collinear, when loop
momenta are parallel to an energetic light quark, and soft, when loop
momenta are much smaller than the $\Lambda_b$ baryon mass $M_{\Lambda_b}$.
Collinear and soft enhancements may overlap to give double logarithms.
Three-particle reducible corrections on the $\Lambda_b$ baryon side are
absorbed into the $\Lambda_b$ baryon wave function. If the light valence
quarks move slowly, collinear divergences associated with these quarks will
not be pinched \cite{LY1}, and soft divergences are important. However,
there is probability, though small, of finding the light quarks in the
$\Lambda_b$ baryon with longitudinal momenta of order $M_{\Lambda_b}$.
Therefore, reducible corrections on the $\Lambda_b$ baryon side are
dominated by soft dynamics, but contain weak double logarithms with 
collinear ones suppressed. Similarly, three-particle reducible corrections
on the $\Lambda_c$ baryon side are absorbed into the $\Lambda_c$ baryon
wave function. In the fast recoil region collinear divergences become
stronger, and double logarithms associated with the $\Lambda_c$ baryon
wave function are more important. The remaining part of radiative 
corrections, with all collinear and soft divergences subtracted, are 
characterized by a scale of order $M_{\Lambda_b}$, and absorbed into the 
hard $b$ quark decay amplitudes. Irreducible corrections, with a gluon 
attaching a quark in the $\Lambda_b$ baryon and a quark in the 
$\Lambda_c$ baryon, are infrared finite in the large recoil region
\cite{LT}, and also absorbed into the hard decay amplitudes.

The kinematic variables are defined as follows. The $\Lambda_b$ baryon is
assumed to be at rest with the momentum
\bea
p\equiv (p^+,p^-,{\bf p}_T)=\frac{M_{\Lambda_b}}{\sqrt{2}}(1,1,{\bf 0})\;.
\eea
The valence quark momenta in the $\Lambda_b$ baryon are parametrized as
\bea
k_1 = (p^+, x_1 p^-, k_{1T})\;,\;\;\;\;
k_2 = (0 , x_2 p^-, k_{2T})\;,\;\;\;\;
k_3 = (0 , x_3 p^-, k_{3T})\;,
\eea
where $k_1$ is associated with the $b$ quark. The momentum fractions and the
transverse momenta obey the conservation laws,
\bea
x_1+x_2+x_3 = 1\;,\;\;\;\; k_{1T}+k_{2T}+k_{3T} = 0\;.
\label{conl}
\eea
The $\Lambda_c$ baryon momentum is chosen as
$p^\prime\equiv ({p^\prime}^+, {p^\prime}^-, {\bf 0})$ with
${p^\prime}^+\gg {p^\prime}^-$ at fast recoil. We define the velocity
transfer $\rho$,
\bea 
\rho = \frac{p\cdot p^\prime}{M_{\Lambda_b} M_{\Lambda_c}} \;,\;\;\;\;
1<\rho< \frac{M_{\Lambda_b}^2+M_{\Lambda_c}^2}
{2 M_{\Lambda_b} M_{\Lambda_c}}\;,
\eea
$M_{\Lambda_c}$ being the $\Lambda_c$ baryon mass. Using the on-shell
condition ${p^\prime}^2=M_{\Lambda_c}^2$, the plus and minus components of
$p^\prime$ are written as
\bea
{p^\prime}^+= \rho_{+} p^+\;,\;\;\;\;
{p^\prime}^-= \rho_{-} p^-\;,
\eea
with
\bea
\rho_{+} = (\rho+\sqrt{\rho^2-1})r\;,\;\;\;\;
\rho_{-} = (\rho-\sqrt{\rho^2-1})r\;,
\eea
and $r=M_{\Lambda_c}/M_{\Lambda_b}$.
The valence quark momenta in the $\Lambda_c$ baryon are parametrized as
\bea
k_1^\prime = (x_1^\prime {p^\prime}^+, {p^\prime}^-, k_{1T}^\prime)\;,
\;\;\;\;
k_2^\prime = (x_2^\prime {p^\prime}^+, 0, k_{2T}^\prime)\;,\;\;\;\;
k_3^\prime = (x_3^\prime {p^\prime}^+, 0, k_{3T}^\prime)\;,
\eea
where $k_1^\prime$ is associated with the $c$ quark.
The primed variables obey similar relations to Eq.~(\ref{conl}).

According to factorization theorem, the hadronic matrix element is
expressed as
\bea
{\cal M_\mu} &=& \int_0^1 [dx] [dx^\prime] \int [d^2k_T] [d^2k^\prime_T]
\bar{\Psi}_{\Lambda_c\alpha'\beta'\gamma'}(k_i^\prime,\mu)
H_{\mu}^{\alpha'\beta'\gamma'\alpha\beta\gamma}
(k_i^\prime,k_i,\rho,M_{\Lambda_b},\mu)
\nonumber\\
& &\times \Psi_{\Lambda_b\alpha\beta\gamma}(k_i,\mu)\;,
\label{fac}
\eea
with the notations
\bea
[dx] = dx_1 dx_2 dx_3 \delta\left(1-\sum_{i=1}^3 x_i\right)\;,\;\;\;\;
[d^2k_T] = d^2k_{1T} d^2k_{2T} d^2k_{3T} \delta^2\left(\sum_{i=1}^3 k_{iT}
\right)\;.
\eea
$[dx^\prime]$ and $[d^2k^{\prime}_T]$ associated with the $\Lambda_c$
baryon are defined in a similar way. The hard amplitude $H_\mu$ will be
computed in Sec.~III. The dependence on the factorization (renormalization)
scale $\mu$ will disappear after performing a RG analysis.

The structure of the $\Lambda_b$ baryon distribution amplitude
$\Psi_{\Lambda_b\alpha\beta\gamma}$ is simplified under the assumptions that
the spin and orbital degrees of freedom of the light quark system are
decoupled, and that the $\Lambda_b$ baryon is in the ground state ($s$-wave).
The distribution amplitude is then expressed as \cite{RA}
\bea
\Psi_{\Lambda_b\alpha\beta\gamma}(k_i,\mu) & = &
\frac{1}{2\sqrt{2}N_c} \int \prod_{l=1}^2 \frac{dy^-_ld{\bf y}_l}
{(2\pi)^3}
e^{ik_l\cdot y_l} \epsilon^{abc} \langle 0|T[b_\alpha^a(y_1)u_\beta^b(y_2)
d_\gamma^c(0)]|\Lambda_b(p)\rangle
\nonumber \\
& = & \frac{f_{\Lambda_b}}{8\sqrt{2}N_c}
[(\sla{p}+M_{\Lambda_b})\gamma_5 C ]_{\beta\gamma} 
[\Lambda_b(p)]_\alpha\Phi (k_i,\mu)\;,
\label{psii}
\eea
where $N_c=3$ is number of colors, $b$, $u$, and $d$ are quark fields, $a$,
$b$, and $c$ are color indices, $\alpha$, $\beta$, and $\gamma$ are spinor
indices, $f_{\Lambda_b}$ is a normalization constant, $C$ is the charge
conjugation matrix, and $\Phi$ is the $\Lambda_b$ baryon wave function. Under
similar assumptions, the $\Lambda_c$ baryon distribution amplitude
$\Psi_{\Lambda_c\alpha\beta\gamma}$ is written as
\bea
\Psi_{\Lambda_c\alpha\beta\gamma}(k'_i,\mu) & = &
\frac{1}{2\sqrt{2}N_c} \int \prod_{l=1}^2 
\frac{d{\yp}^-_l d {\bf {\yp}}_l}{(2\pi)^3}
e^{i\kp_l\cdot \yp_l} \epsilon^{abc}
\langle 0|T[c_\alpha^a(\yp_1)u_\beta^b(\yp_2)
d_\gamma^c(0)]|\Lambda_c(p^\prime)\rangle
\nonumber \\
& = &
\frac{f_{\Lambda_c}}{8\sqrt{2}N_c}
[ (\sla{p}^\prime+M_{\Lambda_c})\gamma_5 C ]_{\beta\gamma} 
[\Lambda_c(p^\prime)]_\alpha\Pi (k_i^\prime,\mu)\;,
\label{psif}
\eea
where the normalization constant $f_{\Lambda_c}$ and the wave function
$\Pi$ are associated with the $\Lambda_c$ baryon.

Because of the inclusion of parton transverse momenta, Sudakov resummation
for a hadron wave function should be performed in the impact parameter $b$
space with $b$ conjugate to $k_T$ \cite{LY1,CS}. The result is \cite{SLL}
\bea
\Phi(k_i^-,b_i,\mu) =
\exp\left[-\sum_{l=2}^3 s(w,k_l^-)
-3 \int_{w}^\mu \frac{d\bar{\mu}}{\bar{\mu}}\gamma_q(\alpha_s(\bar{\mu}))
\right]\phi(x_i)\;,
\label{phiw}
\eea
where $\gamma_q=-\alpha_s/\pi$ is the quark anomalous dimension, and
the factorization scale $w$ is chosen as
\bea
w = \min\left( \frac{1}{{b_1}}, \frac{1}{{b_2}}, 
\frac{1}{{b_3}} \right)\;,
\eea
with $b_3=|{\bf b}_1-{\bf b}_2|$. The explicit expression of the Sudakov
exponent $s$ is given by \cite{BS}
\begin{equation}
s(w,Q)=\int_{w}^{Q}\frac{d p}{p}\left[\ln\left(\frac{Q}
{p}\right)A(\alpha_s(p))+B(\alpha_s(p))\right]\;,
\label{fsl}
\end{equation}
where the anomalous dimensions $A$ to two loops and $B$ to one loop are
\begin{eqnarray}
A&=&C_F\frac{\alpha_s}{\pi}+\left[\frac{67}{9}-\frac{\pi^2}{3}
-\frac{10}{27}n_f+\frac{8}{3}\beta_0\ln\left(\frac{e^{\gamma_E}}{2}\right)
\right]\left(\frac{\alpha_s}{\pi}\right)^2\;,
\nonumber \\
B&=&\frac{2}{3}\frac{\alpha_s}{\pi}\ln\left(\frac{e^{2\gamma_E-1}}
{2}\right)\;,
\end{eqnarray}
$C_F=4/3$ being a color factor, $n_f=4$ the flavor number, and
$\gamma_E$ the Euler constant. The one-loop running coupling constant,
\begin{equation}
\frac{\alpha_s(\mu)}{\pi}=\frac{1}{\beta_0\ln(\mu^2/\Lambda_{\rm QCD}^2)}\;,
\label{ral}
\end{equation}
with the coefficient $\beta_{0}=(33-2n_{f})/12$ and the QCD scale
$\Lambda_{\rm QCD}$, will be substituted into Eq.~(\ref{fsl}). The initial
condition $\phi$ of the Sudakov evolution absorbs nonperturbative dynamics
below the factorization scale $w$.

Following the derivation in \cite{L1,L}, we obtain
the Sudakov resummation for the $\Lambda_c$ baryon distribution
amplitude,
\bea
\Pi(k_i^{\prime +},b_i,\mu) =
\exp\left[-\sum_{l=1}^3 s(w,k_l^{\prime +})
-3 \int_{w}^\mu \frac{d\bar{\mu}}{\bar{\mu}}\gamma_q(\alpha_s(\bar{\mu}))
\right]\pi(x_i^\prime)\;.
\label{piw}
\eea
We have included the Sudakov exponent $s$ associated with the $c$ quark,
which carries large longitudinal momentum in the fast recoil region. Notice
the same transverse extents $b_i$ as those for the $\Lambda_b$ baryon. This
is the consequence of neglecting the transverse momenta which flow through
the virtual quark lines in $H_\mu$ \cite{L}.

The RG analysis of $H_\mu$ leads to
\bea
H_{\mu}(k_i^{\prime +},k_i^-,b_i,\rho,M_{\Lambda_b},\mu) =
\exp\left[-3 \sum_{l=1}^2 \int_\mu^{t_l} \frac{d\bar{\mu}}{\bar{\mu}}
\gamma_q(\alpha_s(\bar{\mu}))\right] 
H_{\mu}(x_i^\prime,x_i,b_i,\rho,M_{\Lambda_b},t_{1},t_{2})\;,
\label{he}
\eea
where the superscripts $\alpha'$, $\beta'$, $\cdots$, have been suppressed.
Since large logarithms have been collected by the exponential, the initial
condition $H_{\mu}$ of the RG evolution on the right-hand side of the above
expression can be computed reliably in perturbation theory.  To simplify the
formalism, we shall make the approximations $M_b\approx M_{\Lambda_b}$ and
$M_c\approx M_{\Lambda_c}$, and neglect the transverse momentum dependence
of the virtual quark propagators as mentioned before. The two arguments
$t_{1}$ and $t_{2}$ of $H_{\mu}$, which will be specified in the next
section, imply that each running coupling constant $\alpha_s$ is evaluated
at the mass scale of the corresponding hard gluon.
Substituting Eqs.~(\ref{psii})-(\ref{he}) into Eq.~(\ref{fac}), we
derive the factorization formula for the semileptonic decay
$\Lambda_b\to \Lambda_cl\bar{\nu}$, where the $\mu$ dependence has
disappeared as stated before.

For the $\Lambda_b$ baryon wave function $\phi(x_1,x_2,x_3)$, we adopt the
model proposed in \cite{S},
\bea
\phi(\zeta,\eta) = N \eta^2\zeta (1-\eta)(1-\zeta) 
\exp\left[-\frac{M_b^2}{2\beta^2(1-\eta)}
-\frac{m_l^2}{2\beta^2\eta\zeta(1-\zeta)}\right]\;,
\eea
with $N$ being a normalization constant, $\beta$ a shape parameter, $m_l$
the mass of light degrees of freedom in the $\Lambda_b$ baryon. The new
variables $\zeta$ and $\eta$ are defined by
\bea
\zeta = \frac{x_2}{x_2+x_3}\;,\;\;\;\;\eta = x_2+x_3 \;.
\eea
In terms of $\zeta$ and $\eta$, the normalization of $\phi(\zeta,\eta)$ is
given by
\bea
\int d\zeta \eta d\eta \phi (\zeta,\eta) = 1 \;,
\label{norma}
\eea
which determines the constant $N$, once the parameters $\beta$ and $m_l$
are fixed. The above wave function with the factor
$\eta^2\zeta (1-\eta)(1-\zeta)=x_1x_2x_3$ suppresses contributions from the
end points of momentum fractions. The exponents proportional to
$M_b^2/(1-\eta)=M_b^2/x_1$ and to
$m_l^2/[\eta\zeta(1-\zeta)]=m_l^2/x_2+m_l^2/x_3$ with $M_b\gg m_l$ indicate
that $\phi$ has a maximum at large $x_1$ and at small $x_2$ and $x_3$, and
that the $b$ quark momentum $k_1^2$ is roughly equal to $M_b^2$.
For $\phi(x_3,x_1,x_2)$ which will appear in the factorization formulas
presented in Sec.~III, the above expression is transformed into
\bea
\phi(\zeta,\eta) = N \eta^2\zeta (1-\eta)(1-\zeta) 
\exp\left[-\frac{M_b^2}{2\beta^2\eta(1-\zeta)}
-\frac{m_l^2(1-\eta+\eta\zeta)}{2\beta^2\eta\zeta (1-\eta)}\right]\;.
\eea
For convenience, we assume that the $\Lambda_c$ wave function
$\pi(\zeta',\eta')$ possesses the same functional form and the same
parameters $\beta$ and $m_l$ as of $\phi(\zeta,\eta)$, but with the $b$
quark mass $M_b$ replaced by the $c$ quark mass $M_c$. The wave function
$\pi(x'_1,x'_2,x'_3)$ also has a maximum at large $x'_1$, such that
the $c$ quark momentum $k_1^{'2}$ is roughly equal to $M_c^2$.

\section{TRANSITION FORM FACTORS}

In this section we present the factorization formulas for the form factors
$f_1$ and $g_1$, which are associated with the spin structures
$\bar{\Lambda}_c\gamma_\mu\Lambda_b$ and $\bar{\Lambda}_c\gamma_\mu\gamma_5
\Lambda_b$ in $M_\mu$, respectively. Working out the contraction of
$\bar{\Psi}_{\Lambda_c\alpha^\prime\beta^\prime\gamma^\prime}
H_{\mu}^{\alpha^\prime\beta^\prime\gamma^\prime\alpha\beta\gamma}
\Psi_{\Lambda_b\alpha\beta\gamma}$ in momentum space, we extract the
hard part $H$. Employing a series of permutations of the valence quark
kinematic variables as in \cite{SLL}, the summation over the leading
diagrams in Fig.~1 reduces to two terms for each form factor. The
factorization formula for the form factors $f_1(\rho)$ and $g_1(\rho)$ are
written as
\bea
f_1(\rho) &=& \frac{4\pi}{27}
\int_0^1[dx^\prime][dx]\int_0^\infty b_1db_1b_2db_2\int_0^{2\pi}d\theta
f_{\Lambda_c}f_{\Lambda_b}
\nonumber \\
& &\times\sum_{j=1}^2 H_j(x^\prime_i,x_i,b_i,\rho,M_{\Lambda_b},t_{jl})
{\cal F}_j(x^\prime_i,x_i,\rho)
\exp[-S(x_i^\prime,x_i,w,\rho,M_{\Lambda_b},t_{jl})] \;,
\label{f1f}\\
g_1(\rho) &=& \frac{4\pi}{27}
\int_0^1[dx^\prime][dx]\int_0^\infty b_1db_1b_2db_2\int_0^{2\pi}d\theta
f_{\Lambda_c}f_{\Lambda_b}
\nonumber \\
& &\times \sum_{j=1}^2 H_j(x^\prime_i,x_i,b_i,\rho,M_{\Lambda_b},t_{jl})
{\cal G}_j(x^\prime_i,x_i,\rho)
\exp[-S(x_i^\prime,x_i,w,\rho,M_{\Lambda_b},t_{jl})] \;,
\label{g1f}
\eea
where $\theta$ is the angle between ${\bf b}_1$ and ${\bf b}_2$.

The functions ${\cal F}_j$ and ${\cal G}_j$, which group together the
products of the initial and final baryon wave functions, are, in terms of
the notations,
\begin{eqnarray}
\phi_{123}\equiv \phi(x_1,x_2,x_3)\;,\;\;\;\;
\pi_{123}\equiv \pi(x_1',x_2',x_3')\;,
\end{eqnarray}
given by
\bea
\frac{{\cal F}_1}{\phi_{123}\pi_{123}}&=&
\frac{r^2}{[(1-\xp_1-\rhom)\rhop+\td](1-\xp_1)x_2\rhop}
\left[2\left(2\srho-1\right)(1-\xp_1) \right.
\nonumber \\
& & \left.+(2(1+r)\rho-4r-1)x_2
+\left(2(2\rho-1)+(2\rho-3)\rho_1\right)x_2\xp_1\right]
\nonumber \\
& & +\frac{r^2}{[(1-\xp_1-\rhom)\rhop+\td](1-x_1)\xp_2\rhop}
\left[\left(\rho_1+2r\sqrt{\rhod-1}+3+4r-r\rho\right)(1-x_1)\right.
\nonumber \\
& & \left.-(\rho_1+3)(1-x_1)\xp_1
   +2\left(2(\rho-1)(\sqrt{\rho+1}+\rho)-1\right)\xp_2\right]
\nonumber \\
& & +\frac{r}{(1-x_1)^2\xp_2\rhop^2}
\left[2r\left(2\srho-1+2\rho\right)(1-x_1)+2\left(\srho-2+\rho\right)\xp_2
\right.
\nonumber \\
& & \left.- r\left((2-\rho)\rho_1+1-2\rho\right)(1-x_1)\xp_2\right]
\nonumber \\
 & & +\frac{r}{(1-x_1)(1-\xp_1)x_2\rhop^2} 
\left[2\left(\srho+2-\rho\right)+r(\rho_1+3)(1-x_1)(1-\xp_1)\right.
\nonumber \\
 & & \left.+2r\left(2(\rho-1)(\srho+\rho)-1\right)x_2 \right]\;,
\label{psi1}
\\
\frac{{\cal F}_2}{\phi_{312}\pi_{312}}&=&
\frac{r}{[(1-\xp_3-\rhom)\rhop+\td](1-x_3)\xp_1\rhop}
\left[ 2r\rho_1(1-\xp_3)+4r^2(1+\rho)(1-x_1)\right.
\nonumber \\
& &+2r^2\left(3 -\srho\right)x_1
-2r\left((\rho-1)\sqrt{\rhod-1}-\rhod\right)\xp_2
\nonumber\\
& &\left.
-(1+\rho_1)(r(\rho-1)x_1+\xp_2)(1-\xp_3) \right]
\nonumber \\
& & +\frac{2r}{[(\xp_2-\rhom)(1-x_1)\rhop+\td][1-(1-x_1\rhop)(1-\xp_2)]}
\nonumber\\
& &\times\left[r\left(\rho+\srho\right)(x_1\xp_2-\rho_1(x_1+\xp_2))
+\rho_1\left(2\td x_1+\xp_2+2r\srho\right) \right]
\nonumber \\
& & +\frac{r}{[1-(1-x_2\rhop)(1-\xp_1)](1-x_3)\rhop}
\left[ 2r\rho_1(1-x_3)+4(1+\rho)(1-\xp_1)
\right.
\nonumber \\
& & -2\left(\sqrt{\rhod-1}-3\right)\xp_1
-2r\left((\rho-1)(\rho+\srho)+1\right)x_2
\nonumber\\
& &\left.
-r(1+\rho_2)\left(rx_2+\srho \xp_1\right)(1-x_3) \right]\;.
\label{psi2}
\\
\frac{{\cal G}_1}{\phi_{123}\pi_{123}}&=&
\frac{r^2}{[(1-\xp_1-\rhom)\rhop+\td](1-\xp_1)x_2\rhop}
\left[(2\rho-3+(2\rho-1)\rho_2)x_2(1-\xp_1) \right.
\nonumber \\
& &  \left.+2\left(2-\rho-\srho\right)x_2
-2\left(2\rho-1+2\srho\right)(1-\xp_1) \right]
\nonumber \\
& & +\frac{r^2}{[(1-\xp_1-\rhom)\rhop+\td](1-x_1)\xp_2\rhop}
\left[2\left(2(\rho-1)(\srho+\rho)-1\right)\xp_2\right.
\nonumber \\
& & \left.+2r\left(\srho+1\right)(1-x_1)-(3\rho_2+1)(1-x_1)(1-\xp_1) \right]
\nonumber \\
& & +\frac{r^2}{(1-x_1)^2\xp_2\rhop^2} 
\left[(2\rho-3+(2\rho-1)\rho_2)\xp_2(1-x_1)
+2\left(2-\rho-\srho\right)\xp_2\right.
\nonumber \\
& &  \left.-2\left(2\rho-1+2\srho\right)(1-x_1) \right]
\nonumber \\
& & +\frac{r}{(1-x_1)(1-\xp_1)x_2\rhop^2} 
\left[-2r\left(2(\rho-1)(\srho+\rho)-1\right)x_2\right.
\nonumber \\
& & \left.-2(\srho+2-\rho)(1-\xp_1)+r(3\rho_2+1)(1-x_1)(1-\xp_1) \right]\;,
\label{gsi1}
\\
\frac{{\cal G}_2}{\phi_{312}\pi_{312}}&=&
\frac{r}{[(1-\xp_3-\rhom)\rhop+\td](1-x_3)\xp_1\rhop}
\left[ -4r^2(1+\rho)+r\rho_2(4-\rho-\rhod)x_1
\right.
\nonumber \\
& & +r(\rho-1)(\rho_2-1)\xp_3x_1
+2r(1-\xp_3)+2r\left((\rho-1)(\srho+\rho)+1\right)\xp_2
\nonumber \\
& &\left.
-(\rho_2+1)\xp_2(1-\xp_3) \right]
\nonumber \\
& & +\frac{2r}{[(\xp_2-\rhom)(1-x_1)\rhop+\td][1-(1-x_1\rhop)(1-\xp_2)]}
\left[r\left(\srho-2-\rho\right)\right.
\nonumber \\
& &\left. +r^2 x_1+\xp_2
-2r\left(\rho+\srho\right)(1-x_1)(1-\xp_2) \right]
\nonumber \\
& & + \frac{r}{[1-(1-x_2\rhop)(1-\xp_1)](1-x_3)\rhop}
\left[ -4(1+\rho)+2r(1-x_3)\right.
\nonumber \\
& &-2\left(\srho+2\rho-1\right)\xp_1
+2r\left((\rho-1)(\srho+\rho)+1\right)x_2
\nonumber\\
& &\left.
    -r(\rho_2+1)(r x_2+(\rho-1)\xp_1)(1-x_3) \right]\;,
\label{gsi2}
\eea
with $\rho_1=\sqrt{(\rho+1)/(\rho-1)}$ and $\rho_2=1/\rho_1$.

The hard parts are given by
\bea
H_1 &=& \alpha_s(t_{11}) \alpha_s(t_{12}) 
K_0\left(\sqrt{(1-x_1)(1-x_1^\prime)\rho_+} M_{\Lambda_b}b_1\right)
K_0\left(\sqrt{x_2x_2^\prime\rho_+} M_{\Lambda_b}b_2\right)\;,
\\
H_2 &=& \alpha_s(t_{21}) \alpha_s(t_{22}) 
K_0\left(\sqrt{x_1x_1^\prime\rho_+} M_{\Lambda_b}b_1\right)
K_0\left(\sqrt{x_2x_2^\prime\rho_+} M_{\Lambda_b}b_2\right) \;,
\eea
with $K_0$ being the modified Bessel function of order zero.
The complete Sudakov exponent $S$ is written as
\bea
S(x_i^\prime,x_i,w,\rho,M_{\Lambda_b},t_{jl})
=S_d(x_i^\prime,x_i,w,\rho,M_{\Lambda_b})+S_s(w,t_{jl})\;,
\eea
with
\bea
S_d &=& \sum_{l=2}^3 s(w,x_l p^-) 
               + \sum_{l=1}^3 s(w,\xp_l {p^\prime}^+)\;,
\\
S_s &=& 3 \int_{w}^{t_{j1}}
\frac{d\bar{\mu}}{\bar{\mu}} \gamma_q(\alpha_s(\bar{\mu}))
+ 3 \int_{w}^{t_{j2}}
\frac{d\bar{\mu}}{\bar{\mu}} \gamma_q(\alpha_s(\bar{\mu}))\;.
\eea
The hard scales $t_{jl}$ are chosen as
\bea
t_{11} &=& \max \left[\sqrt{(1-x_1)(1-x_1^\prime)\rho_+}M_{\Lambda_b},
1/b_1\right]\;,
\nonumber \\
t_{21} &=& \max \left[\sqrt{x_1 x_1^\prime \rho_+}M_{\Lambda_b},
1/b_1\right]\;,
\nonumber \\
t_{21} &=& t_{22} = 
\max \left[\sqrt{x_2 x_2^\prime \rho_+}M_{\Lambda_b},1/b_2\right]\;,
\eea
which are always greater than $w$. It is possible that the hard scales
$t_{jl}$ are small and the running coupling constants become large as $b_i$
are close to $1/\Lambda_{\rm QCD}$. These nonperturbative enhancements are,
however, suppressed by the Sudakov exponential $\exp(-S_d)$, which
decreases quickly in the large $b_i$ region and vanishes as
$b_i \ge 1/\Lambda_{\rm QCD}$. The exponential $\exp(-S_d)$ approaches
unity; that is, there is no Sudakov suppression from the all-order
summation of infrared logarithmic corrections at small $b_i$. In these
short-distance regions higher-order corrections are regarded as being hard
and should be absorbed into $H$ \cite{LS}. Another exponential
$\exp(-S_s)$, as a consequence of single-logarithm summation, describes
the RG evolution from the factorization scale $w$ to the hard scales
$t_{jl}$.

For the case with massless leptons, it is easy to show that the differential
decay rate in the rest frame of the $\Lambda_b$ baryon is given by
\bea
\frac{d\Gamma}{d\rho} &=& \frac{M_{\Lambda_b}^5 r^3}{24 \pi} G_F^2
|V_{cb}|^2\sqrt{\rho^2-1} \{ |f_1|^2 (\rho-1) [3+3 r^2 -2 (2 \rho-1) r] 
\nonumber\\
& &+ |g_1|^2 (\rho+1) [3+3 r^2-2 (2 \rho+1) r] \}\;,
\label{decay}
\eea
where only the contributions from the form factors $f_1$ and $g_1$ are
considered. It is straightforward to obtain the total decay rate
\begin{equation}
\Gamma\equiv \int d\rho\frac{d\Gamma}{d\rho}
\label{tde}
\end{equation}
from Eq.~(\ref{decay}) and thus the branching ratio
$B(\Lambda_b\rightarrow \Lambda_c l\bar{\nu})$, if the form factors
$f_1(\rho)$ and $g_1(\rho)$ in the whole range of $\rho$ are known.

\section{RESULTS}

In order to reduce the number of unknown parameters, we make an 
approximation. Consider the baryonic decay constant $\tilde{f}_{\Lambda}$
defined, in heavy quark effective theory, by
\bea
\langle 0|\tilde{j}^v|\Lambda_Q\rangle &=& \tilde{f}_\Lambda \Lambda_Q \;,
\label{bdc}
\eea
in terms of the $\Lambda$-baryonic current \cite{Dai1,Dai2}
\bea
\tilde{j}^v = \epsilon^{abc}(u^a C\gamma_5 d^b)h_v^c\;,
\eea
where $\Lambda_Q$ is the heavy baryon spinor, $h_v$ the heavy
quark field, and $a$, $b$, $c$ denote the color indices.
We contract a Dirac tensor $(C\gamma_5)_{\beta\gamma}$ with
a heavy $\Lambda$-baryon distribution amplitude such as
$\Psi_{\Lambda_b\alpha\beta\gamma}$ in Eq.~(\ref{psii}) and integrate out
the valence quark momenta $k_i$. Compared with Eq.~(\ref{bdc}), we
extract the baryonic decay constant 
\bea
\tilde{f}_{\Lambda} = f_{\Lambda_Q} M_{\Lambda_Q} \;.
\eea
It implies that in the heavy quark limit the normalization constants 
$f_{\Lambda_b}$ and $f_{\Lambda_c}$ are related by
\bea
f_{\Lambda_b} M_{\Lambda_b} &=& f_{\Lambda_c} M_{\Lambda_c}\;.
\label{barj}
\eea
Therefore, $f_{\Lambda_c}$ associated with the $\Lambda_c$ baryon will
not be treated as a free parameter in the numerical analysis below.

We are now ready to compute the form factors $f_1(\rho)$ and $g_1(\rho)$
from Eqs.~(\ref{f1f}) and (\ref{g1f}), adopting the CKM matrix element
$V_{cb}=0.04$, the masses $M_{\Lambda_b}=5.624$ GeV and
$M_{\Lambda_c}=2.285$ GeV, and the QCD scale $\Lambda_{\rm QCD}=0.2$ GeV.
We examine the self-consistency of our calculation by considering the
percentage of the full contribution to the form factor $f_1$, that arises
from the short-distance region with all $\alpha_s(t_{jl})/\pi<0.5$. The
percentages for different $\beta$ with $m_l$ fixed at 0.3 GeV are listed
in Table I. It is observed that the perturbative contributions become
dominant gradually as $\rho$ and $\beta$ increase: a larger $\rho$ 
corresponds to larger momentum transfer involved in decay processes, and 
a larger $\beta$ corresponds to heavy baryon wave functions which are 
less sharp at the high ends of the momentum fractions $x_1$ and $x'_1$. 
We conclude that the PQCD analysis of the transition form factors is 
self-consistent for $\beta >1.0$ GeV and $\rho > 1.2$, viewing the 
perturbative percentage of about 80\%. Compared to the corresponding 
meson decay $B\to Dl{\bar\nu}$ \cite{L1}, a perturbative expansion is 
less reliable in the baryon case, because partons in a baryon are softer, 
such that Sudakov suppression is weaker.

To obtain the total decay rate, we need the information of $f_1$ and
$g_1$ in the whole range of $\rho$. Since the perturbative analysis is
reliable only in the fast recoil region, we extrapolate the PQCD
predictions at large $\rho$ to small $\rho$. Hinted by \cite{HYC}, we 
propose the following parametrization for the form factors:
\begin{equation}
f_1(\rho)=\frac{c_f}{\rho^{\alpha_f}}\;,\;\;\;\;
g_1(\rho)=\frac{c_g}{\rho^{\alpha_g}}\;,
\label{par}
\end{equation}
where the constants $c_f$ and $c_g$, and the powers $\alpha_f$ and
$\alpha_g$ are determined by the PQCD results at large $\rho$. The
constants $c_f$ and $c_g$, equal to the values of the form factors at
zero recoil ($\rho=1$), should be close to unity according to heavy
quark symmetry. We fit Eq.~(\ref{par}) to the PQCD results in the range
with $\rho>1.3$ for $\beta=1.0$, where perturbative contribution has
exceeded 80\%. The powers $\alpha_f=5.18$ and $\alpha_g=5.14$, close to
$\alpha_f\sim 4.6$ at large $\rho$ from the method of wave function
overlap integrals \cite{XHG}, are obtained. These values are larger
than 1.8 extracted from the transition form factors associated with the
corresponding meson decay $B\to Dl{\bar\nu}$ \cite{L1}. This is
expected, because perturbative baryon decays involve more hard gluon
exchanges.

On the experimental side, there exist only the data of the semileptonic
branching ratio $B(\Lambda_b\to Xl{\bar\nu})\sim 10\%$ \cite{Data}, where
the final-state particles $X$ are dominated by the charm baryons. The
data of the $B$ meson semileptonic decays show
$B(B\to D^* l\bar{\nu})\sim 3B(B\to D l\bar{\nu})$, indicating that each
of the three polarization states of the $D^*$ meson contributes the 
same amount of branching ratio as the $D$ meson does. It is possible
that this observation applies to dominant modes in the
$\Lambda_b\to Xl{\bar\nu}$ decays with the excited charm baryons
$\Lambda_c(2593)$ of spin $J=1/2$ and $\Lambda_c(2625)$ of $J=3/2$.
That is, the branching ratio $B(\Lambda_b\to \Lambda_cl{\bar\nu})$ is
about $1/4$ of $B(\Lambda_b\to Xl{\bar\nu})$, {\it i.e.}, about
$2\sim 3\%$. This estimation is consistent with the experimental
upper bound of the branching ratio from the data
$B(\Lambda_b\to \Lambda_cl{\bar\nu}+X)=(8.27\pm 3.38)\%$ \cite{Data}.

We substitute Eq.~(\ref{par}) for the form factors $f_1$ and $g_1$ into the
decay rate $\Gamma$ in Eq.~(\ref{tde}), and adjust the normalization
constant $f_{\Lambda_b}$ such that our predictions for the branching ratio
are located in the range of $2\sim 3\%$. The
$\Lambda_c$ baryon normalization constant $f_{\Lambda_c}$ changes according
to Eq.~(\ref{barj}).  We adopt the $\Lambda_b$ baryon lifetime
$\tau=(1.24\pm 0.08)\times 10^{-12}$ s\cite{Data}. The value of 
$f_{\Lambda_b}$ determines the parameters $c_f$ and $c_g$. It is then found
that $f_{\Lambda_b}=2.71 \times 10^{-3}$ GeV$^2$,
corresponding to
\begin{equation}
f_1(\rho)=\frac{1.32}{\rho^{5.18}}\;,\;\;\;\;
g_1(\rho)=\frac{- 1.19}{\rho^{5.14}}\;,
\label{fit1}
\end{equation}
gives the branching ratio 2\%, and $f_{\Lambda_b}=3.0 \times 10^{-3}$
GeV$^2$, corresponding to
\begin{equation}
f_1(\rho)=\frac{1.62}{\rho^{5.18}}\;,\;\;\;\;
g_1(\rho)=\frac{- 1.46}{\rho^{5.14}}\;,
\end{equation}
gives the branching ratio 3\%. Since the values of the form factors at zero
recoil should be close to unity as stated above, we prefer 
Eq.~(\ref{fit1}) with $f_1(1)=1.32$ and $g_1(1)=-1.19$, which are
also consistent with the conclusion in \cite{XHG}. The corresponding
normalization constant $f_{\Lambda_b}=2.71\times 10^{-3}$ GeV$^2$, of the
same order as $f_{P}=(5.2\pm 0.3)\times 10^{-3}$ GeV$^2$ for the proton
\cite{CZ1}, is reasonable. The PQCD predictions and the corresponding
extrapolations are displayed in Fig.~2, which deviate from each other at
small $\rho$. If applying the PQCD formalism to the zero recoil region, we
shall obtain divergent form factors as shown in Fig.~2, which imply the
failure of PQCD. Note that our results of the form factors exhibit slopes
larger than the dipole behavior assumed in \cite{HYC}.

We then examine the sensitivity of our predictions for the branching
ratio $B(\Lambda_b\to \Lambda_cl{\bar\nu})$ to the variation of the
parameter $\beta$. Choosing $\beta=2.0$ GeV and $\beta=4.0$ GeV, and
normalizing the corresponding form factors in the way that they have 
similar values to those for $\beta=1.0$ GeV in Eq.~(\ref{fit1}),
we obtain the form factors
\begin{equation}
f_1(\rho)=\frac{1.34}{\rho^{5.04}}\;,\;\;\;\;
g_1(\rho)=\frac{- 1.17}{\rho^{4.92}}\;,
\label{fit3}
\end{equation}
and
\begin{equation}
f_1(\rho)=\frac{1.34}{\rho^{4.94}}\;,\;\;\;\;
g_1(\rho)=\frac{- 1.18}{\rho^{4.79}}\;,
\label{fit4}
\end{equation}
respectively. Equations (\ref{fit3}) and (\ref{fit4}) lead to increases 
of the branching ratio by 4\% and 8\%, respectively. That is, our 
predictions for the branching ratio are not sensitive to the choice of 
baryon wave functions. This observation is attributed to the fact that 
the PQCD results of the transition form factors at large 
recoil are insensitive to the variation of baryon wave functions.

We present in Fig.~3 the differential decay rate $d\Gamma/d\rho$
derived from the form factors in Eq.~(\ref{fit1}), which
can be compared with experimental data in the future. The $\Lambda_b$ and
$\Lambda_c$ baryon wave functions determined in this work are given by
\bea
\phi(\zeta,\eta) &=& 6.67 \times 10^{12}\; \eta^2\zeta (1-\eta)(1-\zeta)
\nonumber\\
& &\times 
\exp\left[-\frac{M_b^2}{2(1.0\;\;{\rm GeV})^2(1-\eta)}
-\frac{m_l^2}{2(1.0\;\;{\rm GeV})^2\eta\zeta(1-\zeta)}\right]\;,
\label{bw}\\
\pi(\zeta,\eta) &=& 6.94\times 10^{4}\; \eta^2\zeta (1-\eta)(1-\zeta) 
\nonumber\\
& &\times
\exp\left[-\frac{M_c^2}{2(1.0\;\;{\rm GeV})^2(1-\eta)}
-\frac{m_l^2}{2(1.0\;\;{\rm GeV})^2\eta\zeta(1-\zeta)}\right]\;.
\label{pw}
\eea

At last, we compare our predictions with those derived from other
approaches in the literature. The $\Lambda_b\to\Lambda_c$ transition
form factors have been evaluated by means of overlap integrals of
infinite-momentum-frame (IMF) wave functions, nonrelativistic and
relativistic quark models, and QCD sum rules. For a review, refer to
\cite{KKP}. Basically, they are nonperturbative methods without involving
hard gluons. QCD dynamics is completely parametrized into IMF wave
functions in the overlap-integral approach \cite{XHG,KKK}, and into
baryon-three-quark vertex form factors in the relativistic quark model
\cite{IKL1}. Information of the above bound-state quantities can be
obtained by solving Bethe-Salpeter equations \cite{IKL2}. Most of the
analyses, including QCD sum rules \cite{Dai2,DFN,CNN}, led to
the branching ratios about or below 6\%. The prediction
$B(\Lambda_b\to\Lambda_c l{\bar\nu})\sim 9\%$ in \cite{KKK} is a bit
higher compared to the data of $B(\Lambda_b\to \Lambda_cl{\bar\nu}+X)$. 
Our result is close to $(3.4\pm 0.6)\%$ derived in \cite{DFN}.

\section{CONCLUSION}

In this paper we have developed PQCD factorization theorem for the
semileptonic heavy baryon decay $\Lambda_b \to \Lambda_c l\bar{\nu}$, whose 
form factors are expressed as the convolutions of hard $b$ quark decay
amplitudes with universal $\Lambda_b$ and $\Lambda_c$ baryon wave functions.
It is observed that the PQCD formalism with Sudakov suppression in
the long-distance region is applicable to $\Lambda_b \to \Lambda_c$
decays for the velocity transfer greater than 1.2. This observation
indicates that PQCD is an appropriate approach to analyses of two-body
exclusive nonleptonic $\Lambda_b$ baryon decays. Requiring that the 
normalizations of the form factors at zero recoil are consistent with 
heavy quark symmetry, we have predicted the branching ratio
$B(\Lambda_b \to \Lambda_c l{\bar\nu})\sim 2\%$. We have also
determined the $\Lambda_b$ and $\Lambda_c$ baryon wave functions shown
in Eqs.~(\ref{bw}) and (\ref{pw}), respectively.
These wave functions, because of their universality, will be employed
to study nonleptonic $\Lambda_b$ baryon decays in the future.

\vskip 1.0cm
This work was supported by the National Science Council of the Republic of
China under Grant Nos. NSC-88-2112-M-001-041 and NSC-88-2112-M-006-013.
%\vskip 1.0cm

%\vskip 2.0cm
\newpage

\centerline{\large\bf FIGURE CAPTIONS}
\vskip 0.5cm

\noindent
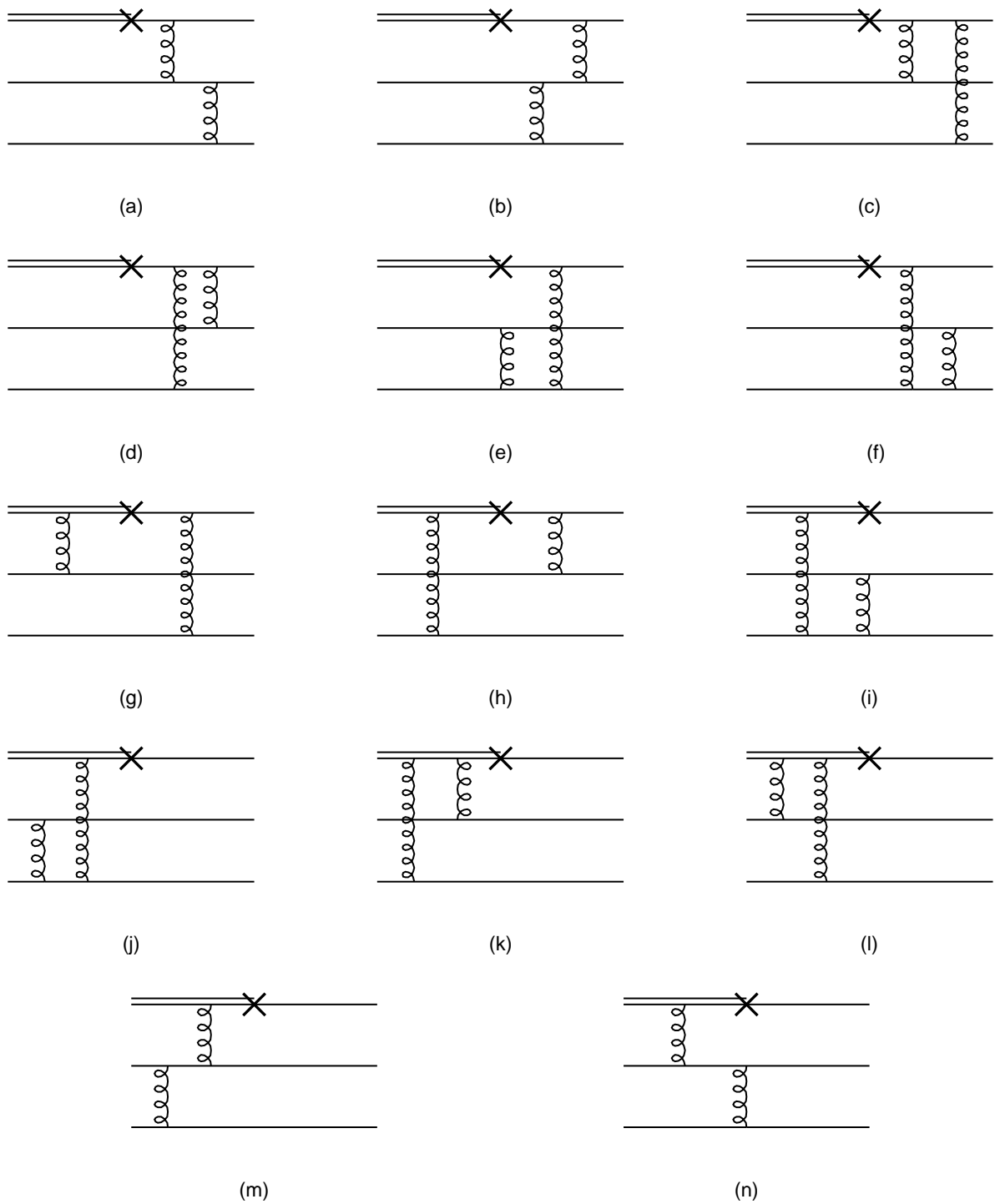
{\bf FIG. 1} Lowest order diagrams for the
$\Lambda_b\to \Lambda_c l\bar{\nu}$ decay.
\vskip 0.3cm

\noindent
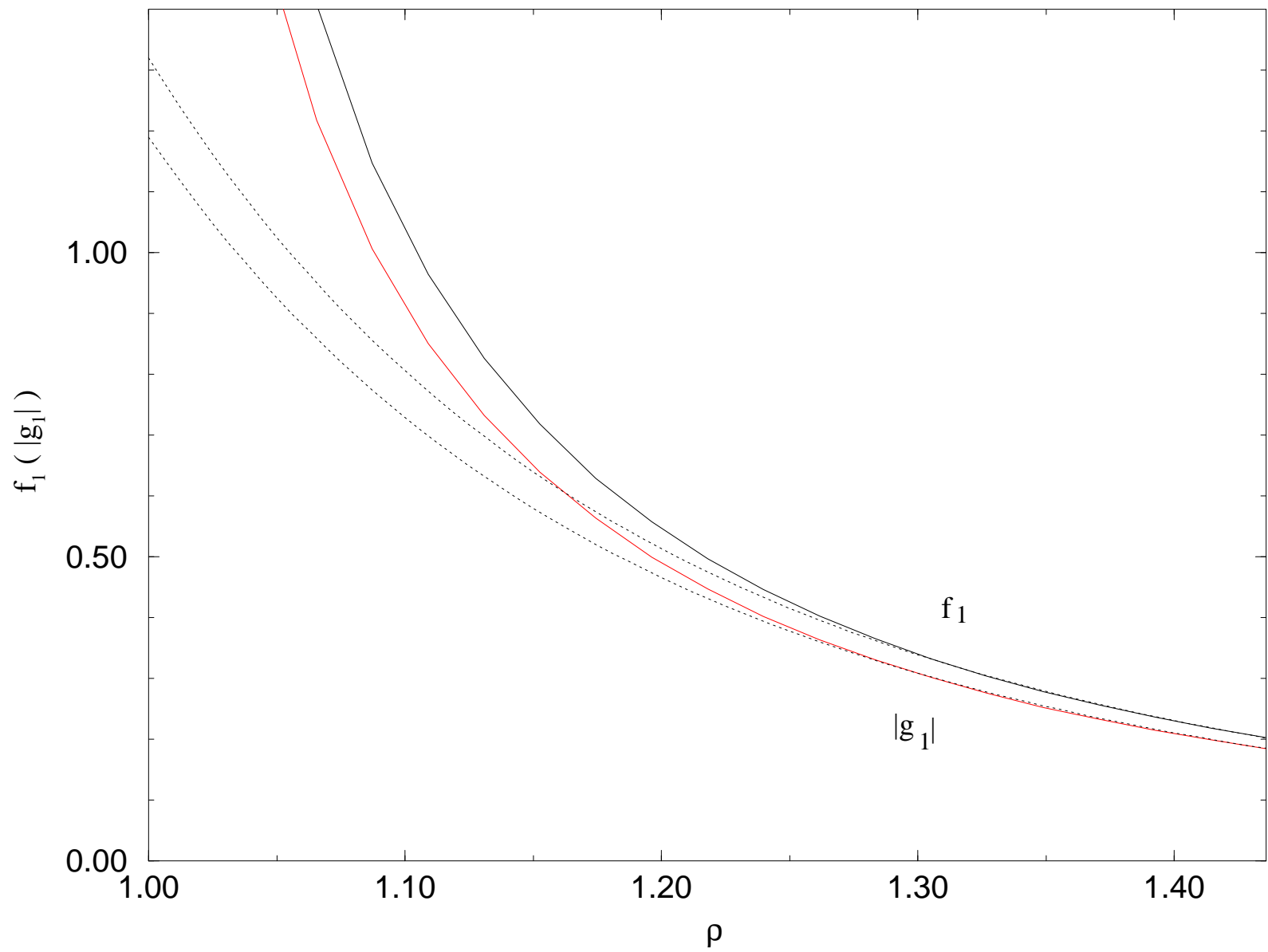
{\bf FIG. 2} Dependence of $f_1$ and $|g_1|$ on $\rho$ for $\beta=1.0$ and
$m_l=0.3$ obtained from PQCD (solid lines) and from the extrapolation
in Eq.~(\ref{fit1}) (dashed lines). The upper (lower) set of curves
represents the form factor $f_1$ ($|g_1|$).
\vskip 0.3cm

\noindent
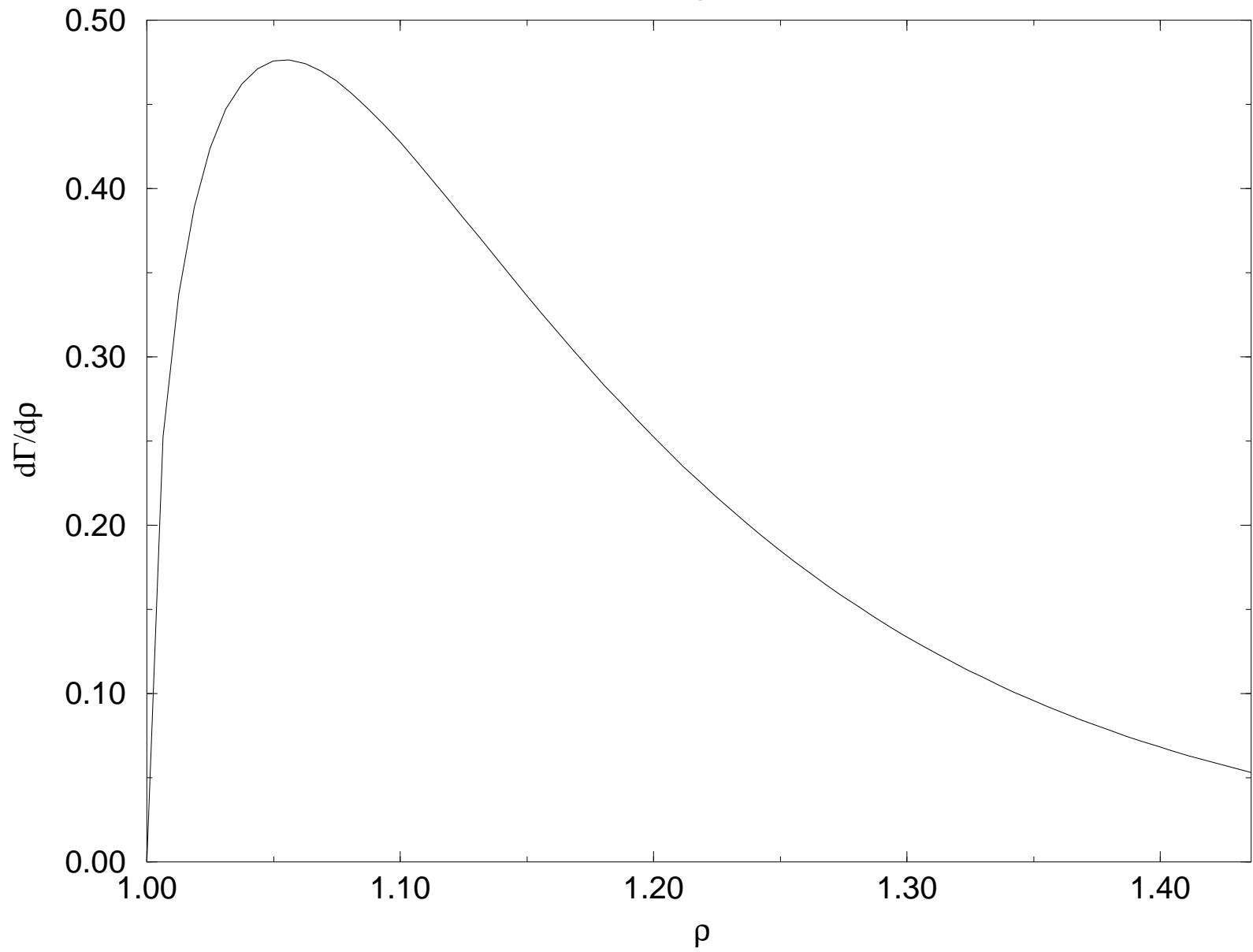
{\bf FIG. 3} Dependence of $d\Gamma/d\rho$ on $\rho$ obtained from
Eq.~(\ref{fit1}) in units of $10^{-13}$ GeV.
\vskip 0.3cm

%\vskip 2.0cm
\newpage
Table I.
Percentages of perturbative contributions for various $\beta$ and
$\rho$.
\vskip 0.5cm
\begin{center}
\begin{tabular}{cccc}
\hline
 percentage       &   $\rho=1.2$  &  $\rho=1.3$   &   $\rho=1.4$ 
\\
\hline
 $\beta=1.0$ GeV  &      77.7\%  &  83.6\% &  85.2\% 
\\
\hline
 $\beta=2.0$ GeV  &      79.3\%  &  83.0\% &   85.7\%
\\
\hline
 $\beta=4.0$ GeV  &      82.3\%  &  84.7\% &   86.3\%
\\
\hline

\end{tabular}
\end{center}


\newpage

\begin{thebibliography}{99}
\bibitem{LY1} H-n. Li and H.L. Yu, Phys. Rev. Lett. {\bf 74}, 4388 (1995);
Phys. Rev. D {\bf 53}, 2480 (1996).
\bibitem{CL} C.H. Chang and H-n. Li, Phys. Rev. D {\bf 55}, 5577 (1997);
T.W. Yeh and H-n. Li, Phys. Rev. D {\bf 56}, 1615 (1997).
\bibitem{L1} H-n. Li, Phys. Rev. D {\bf 52}, 3958 (1995).
\bibitem{BSW} M. Bauer, B. Stech and M. Wirbel, Z. Phys. C {\bf 29}, 637 
(1985); {\bf 34}, 103 (1987).
\bibitem{CLY} H.Y. Cheng, H-n. Li, and K.C. Yang, Phys. Rev. D {\bf 60}, 
094005 (1999).
\bibitem{WYL} C.Y. Wu, T.W. Yeh, and H-n. Li, Phys. Rev. D {\bf 55}, 237
(1997).
\bibitem{C} H.Y. Cheng, Phys. Lett. B {\bf 335}, 428 (1994).
\bibitem{SLL} H.H. Shih, S.C. Lee, and H-n. Li, 
              Phys. Rev. D {\bf 59}, 094014 (1999).
\bibitem{RA} W. Loinaz and R. Akhoury, Phys. Rev. D {\bf 53}, 1416 (1996).
\bibitem{Li1} H-n. Li, hep-ph/9904449.
\bibitem{Li2} W.F. Chang, H-n. Li, and H.L. Yu, Phys. Lett. B {\bf 457}, 
341 (1999).
\bibitem{NS} M. Neubert and C.T. Sachrajda, 
             Nucl. Phys. {\bf B483}, 339 (1997).
\bibitem{Alt} G. Altarelli, G. Martinelli, S. Petrarca and F. Rapuano,
              Phys. Lett. B {\bf 382}, 409 (1996).
\bibitem{HYC1} H.Y. Cheng, Phys. Rev. D {\bf 56}, 2783 (1997).
\bibitem{LT} H-n. Li and B. Tseng, Phys. Rev. D {\bf 57}, 443 (1998)
\bibitem{CS} J.C. Collins and D.E. Soper, Nucl. Phys. {\bf B193}, 381 (1981).
\bibitem{BS} J. Botts and G. Sterman, Nucl. Phys. {\bf B325}, 62 (1989).
\bibitem{L} H-n. Li, Phys. Rev. D {\bf 48}, 4243 (1993);
B. Kundu, H-n. Li, J. Samuelsson, and P. Jain,
Euro. Phys. J. C {\bf 8}, 637 (1999).
\bibitem{S} F. Schlumpf, Ph.D. thesis, 1992, Report No. hep-ph/9211255.
\bibitem{LS} H-n. Li and G. Sterman, Nucl. Phys. {\bf B381}, 129 (1992).
\bibitem{Dai1} Y.B. Dai, C.S. Huang, C. Liu and C.D. L\"{u},
Phys. Lett. B {\bf 371}, 99 (1996).
\bibitem{Dai2} Y.B. Dai, C.S. Huang, M.Q. Huang and C. Liu,
Phys. Lett. B {\bf 387}, 379 (1996).
\bibitem{HYC} H.Y. Cheng and B. Tseng, 
Phys. Rev. D {\bf 53}, 1457 (1996).
\bibitem{XHG} X.H. Guo and P. Kroll, Z. Phys. C {\bf 59}, 567 (1993).
\bibitem{Data} Particle Data Group, C. Caso {\it et al.},
Euro. Phys. J. C {\bf 3}, 1 (1998).
\bibitem{CZ1} V.L. Chernyak and A.R. Zhitnitsky,
Phys. Rep. {\bf 112}, 173 (1984).
\bibitem{KKP} J.G. K\"orner, M. Kr\"amer, and D. Pirjol, Prog. in Part.
Nucl. Phys. {\bf 33}, 787 (1994); M. Neubert, Phys. Rep. {\bf 245}, 259
(1994).
\bibitem{KKK} B. K\"onig, J.G. K\"orner, M. Kr\"amer, and P. Kroll, Phys.
Rev. D {\bf 56}, 4282 (1997).
\bibitem{IKL1} M.A. Ivanov, J.G. K\"orner, V.E. Lyubovitskij, and
A.G. Rusetsky, Report No. hep-ph/9910342 (and references therein).
\bibitem{IKL2} M.A. Ivanov, J.G. K\"orner, V.E. Lyubovitskij, and
A.G. Rusetsky, Phys. Rev. D {\bf 59}, 074016 (1999).
\bibitem{DFN} H.G. Dosch, E. Ferreira, M. Nielsen, and R. Rosenfeld,
Phys. Lett. B {\bf 431}, 173 (1998).
\bibitem{CNN} R.S.M. de Carvalho, F.S. Navarra, M. Nielsen, E. Ferreira,
and H.G. Dosch, Phys. Rev. D {\bf 60}, 034009 (1999).


\end{thebibliography}
\end{document}